\documentclass[aps,prc,twocolumn,twosided]{revtex4}
\usepackage{epsfig,amsmath}

\begin{document}

\title{Dilepton-tagged jets in relativistic nucleus-nucleus collisions: A
case study}
\author{Dinesh K.~Srivastava}
\altaffiliation{Permanent address: Variable Energy Cyclotron Centre, 
             1/AF Bidhan Nagar, Kolkata 700 064, India}            
\affiliation{Physics Department, McGill University,
             3600 University Street, Montreal, H3A 2T8, Canada} 
\author{Charles Gale}
\affiliation{Physics Department, McGill University,
             3600 University Street, Montreal, H3A 2T8, Canada} 
\author{T. C. Awes}
\affiliation{Oak Ridge National Laboratory, Oak Ridge,
Tennessee 37831-6372, USA}
\date{\today}
\begin{abstract}
We study the $A+B \to \ell^+ \ell^- +$ jet +X process in nucleus-nucleus
collisions at relativistic energies. The dilepton as well as the jet
will pass through the matter  produced in
such collisions. The recoiling dilepton will carry information about
the kinematical features of the jet, and will thus prove to be a very effective tool
in isolating in-medium effects  such as energy-loss and fragmentation
function modifications. 
We estimate the contributions due to correlated charm and bottom decay
and we identify a window where they are small as compared to pairs from the 
NLO Drell-Yan process.
\end{abstract}
\maketitle

\section{Introduction}

The study of dilepton production through the 
Drell-Yan process~\cite{dy}, $h_1 + h_2 \to \ell^+ \ell^- +X$ has remained 
a useful tool to sound out concepts about the
parton model, QCD, and the structure functions of hadrons. If the
initial state partons have no transverse momentum, the
lowest-order process $q\overline{q} \to \gamma^* \to \ell^+ \ell^-$
produces a lepton pair with a net $q_T$=0. Experiments, however, show
that the net transverse momentum of the dileptons produced by the Drell-Yan
process are of the order 1 GeV for a dilepton having a mass  $ M \sim $
10 GeV \cite{field}. This fact outlines the importance of 
the NLO Drell-Yan
contributions.   Also, one could assign an
intrinsic spread to the transverse momenta of the partons. This momentum can
have many origins, and a confinement
argument along those lines only accounts for 
$<q_T^2> \sim $ (0.3 GeV)$^2$ \cite{gale}.

It is therefore clear that the dileptons acquire additional
 transverse momenta from production
mechanisms beyond leading order in perturbation theory:
\begin{equation}
q g \to q \gamma^* ~~~~~{\rm {and}}~~~  q \overline{q} \to g \gamma^*~~.
\end{equation}
These Compton and annihilation processes are analogous to the ones
responsible for the 
production of real photons where the recoiling final state
quark or gluon balances the $q_T$ of the dilepton. An important 
point~\cite{kaj} is that the dileptons thus produced 
are {\em always} accompanied by a recoiling quark or gluon.
 If the energy of the
quark or the gluon is several tens of GeV, it will lead to a
jet of hadrons in a narrow cone around the leading hadron in the jet.

Now consider a collision of two heavy nuclei at relativistic energies, 
which could lead to the formation of a quark gluon plasma (QGP). 
The dilepton and the jet
produced in the above process will pass through the plasma. The dileptons
will not interact, while the jet will lose energy through collisions and
radiation in the plasma and provide valuable information about 
these mechanisms in the medium. The dilepton can thus
be used to tag the jet in the same manner that  
photon-tagged jet measurements have been suggested
as a precise probe for the study of jet-quenching~\cite{xnw}.

We recall that jet-quenching can  manifest itself in various ways;
there would be a suppression of hadrons having large transverse
momenta, when compared to the results for $pp$ collisions,
and the fragmentation function in $AA$ collision may be different
from those seen in $e^+e^-$ or $pp$ collisions, etc \cite{xnw}. A precise 
determination
of these effects will require knowledge of the energy of the 
parton produced in the hard
collision which fragmented into the jet. 
Note that the inclusive transverse momentum 
distribution of the hadrons  will not permit us to deduce 
the value for the parton $dE/dx$ unambiguously, as a hadron 
with a given $p_T^{\rm hadron}$ can arise from the fragmentation of any
parton having $p_T^{\rm parton}=p_T^{\rm hadron}/z$, where
$p_T^{\rm parton}$ is the momentum of the parton {\em at the
time of fragmentation}, and $z$ $(<  1)$,
is the fragmentation variable. Recall also that the vast amount of
theoretical activity~\cite{dedx} in the last decade has yielded varying
predictions for the
dependence of $dE/dx$ on the energy of the parton and the properties
of the medium. A precise knowledge of the energy of the progenitor
of the jet of hadrons will provide the information necessary to settle 
these issues.

\begin{figure}[tb]
  \begin{center}
  \epsfig{file=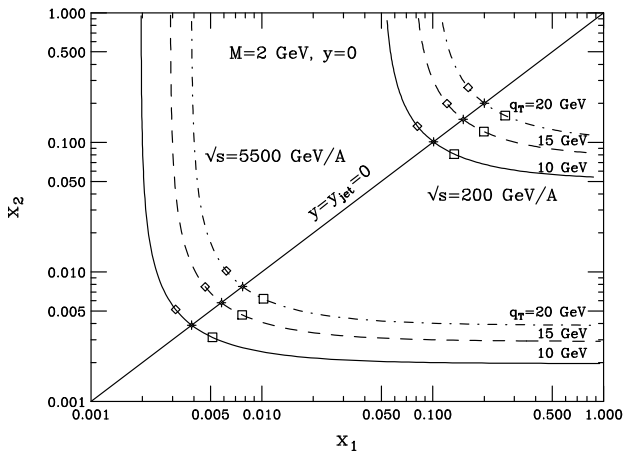,width=8.2cm}
  \caption{The fractional momenta of the partons at RHIC and LHC
energies relevant for the production of dileptons through the NLO
Compton and annihilation processes. The symbols on the curves 
denote the points when $y=0$ and $y_{\rm jet}$ =0.5 (squares), 0.0 (crosses)
and -0.5 (diamonds).}
  \label{fig1}
  \end{center}
\end{figure}

The very large background of hadrons in $AA$ 
collisions makes it very difficult to characterize the jet through 
canonical jet algorithms which typically prescribe energy measurements 
in some cone around the jet-axis.  To minimize contributions from the large 
non-jet background, particles emitted beyond the cone or 
having an energy 
below some threshold are excluded in the jet measurement. This excluded energy,
as well as the non-jet background energy, must be estimated and
taken into account. Because of the difficulty of making these corrections,
photon tagging of the jet is a valuable alternative or complement to the
jet energy measurement, as the 
transverse momentum of the photon (real or virtual) is equal to that 
of the high energy parton produced in the collision.

In this work, a measurement of dilepton-tagged
jets is suggested and given quantitative support. 
We argue that these measurements
could have advantages over measurements of real photons,
even though they are similar in philosophy.
In measurements of real photons, photons from the radiative decays of 
mesons, predominantly
$\pi^0$'s, constitute a huge background which overwhelms the real photon
yield except at very high transverse momenta. The real photon sample can be
enhanced by rejecting photons for which an accompanying photon is measured
with the two photons having a photon-pair mass which falls 
within the $\pi^0$ mass window. However, this method fails at low transverse 
momentum for central heavy ion collisions due to the 
need for a large acceptance in order 
to measure both photons with high probability, 
and due to the high $\pi^0$ multiplicity
and the corresponding large number of combinatorial photon-pairs which results
in the rejection of essentially all photons as possible decay photons.
As the transverse momentum of the $\pi^0$ increases, 
such as when it is the leading hadron in the jet, the 
opening angle between the two decay photons decreases, making it possible
to identify and 
reject the decay photons with high probability and with fewer 
false combinatorial possibilities. However, for photon 
measurements using calorimeters, at large
transverse momentum the opening angle becomes so small that 
the symmetric decay photons will merge into a single shower
cluster and appear as a single isolated photon, 
making it impossible
to isolate them and discriminate between a single real 
photon and two overlapping
photons from the decay of the $\pi^0$.  
This then leaves a narrow window of photon (jet) energies
where the real photons can be identified and such studies
may be conducted. Of course, quantitative limits will depend on actual detector
characteristics of geometry and granularity, and on the particle 
multiplicity. Typical calorimeter geometries have $\pi^0$-photon
shower merging limits as low as 20 GeV/c.

Now consider the case of a dilepton recoiling against the jet. The $\ell^+$ and
$\ell^-$ are easily separated for arbitrarily large momenta. There 
are no other sources of dileptons having large $q_T$ (see later).
Furthermore, the mass of the lepton-pair $M$ gives an 
additional handle on the 
initial state scattering~\cite{xnw},
as we can study dileptons which have the same transverse momentum (i.e.
the recoiling jets have the same transverse momentum) but different
masses. One pays the price of a low counting rate, but if the signal can be
separated from the background and if luminosities are sufficient, we have a
versatile tag.

  To repeat, 
the dileptons produced through the process $q\overline{q} \to \ell^+ \ell^-$
will have only  a modest $q_T$ resulting from the intrinsic transverse momentum
of the partons.
The correlated charm and bottom decays, which offer a huge contribution
to the dilepton production, will also be governed by the same order of $q_T$
given from the intrinsic momenta of the partons, in lowest order.
Even though the $c\overline{c}$ and the $b\overline{b}$ will have a 
vanishing 
$q_T$ in the lowest order, the random transverse momenta of the leptons
in the semi-leptonic decay of the resulting D and B mesons could still
result in a net $q_T$ for the lepton-pair. 
At the NLO, the heavy-quark pairs will have additional  $q_T$,
which would then get translated into the $q_T$ of the
lepton-pair. One can see that the heavy quark will transfer only a part of its
momentum to the lepton, (the semi-leptonic decays B $\to$ D$\ell \nu$
and D $\to$ K$\ell\nu$ have three-body final states) and thus there is
hope that for large $q_T$ of the pair, the contribution of the heavy-quark
decays may become smaller than the NLO Drell-Yan contribution. 
It is also likely that one may be able to suppress or account for these
contributions explicitly if one can identify the vertex of the 
decaying D or B meson, or reject leptons within a region of jet-like 
activity. 
Moreover, if the charm and the bottom quarks lose energy 
due to collisions or radiations in the plasma~\cite{ben}, their contribution
will again be reduced~\cite{shur}, 
as the individual momenta of the quarks will be reduced, 
leading to a reduction in the net $q_T$ of the resulting lepton-pair.

These hopeful considerations are quantitatively tested in the following.
Our goal here is to perform simple phenomenological estimates, and to identify
kinematical domains appropriate to RHIC and to the LHC where our tag would
shine through the background. 
Note in passing that the CMS experiment has considered the
possibility of observing jets tagged by W and Z bosons, which are governed
by the same criteria~\cite{cms_note}, and are thus open to similar
vulnerabilities. See also \cite{kks}.

\section{ Dilepton production}

At the outset, we add that the thermal radiation of dileptons
from the QGP and the hot hadronic matter, a subject of considerable 
research~\cite{gale} will not be important~\cite{ramona} in the domain of
very large $q_T$ of interest here. The dileptons from the
annihilation of jets passing through the QGP~\cite{jet_dil}, could
however have some contribution, at least at the LHC. We shall report
on this in a future publication. In the following we consider
only the Drell-Yan process and  the correlated decay of heavy mesons.

\subsection{ Drell-Yan : Lowest Order }

The  cross section for the lowest-order Drell-Yan process is given by
\begin{equation}
  \frac{d\sigma}{dM^2\,dy} = \frac{4 \pi \alpha^2}{9M^4}\, 
F(x_1,x_2,M^2)
\label{lody}
\end{equation}
where
\begin{eqnarray}
F(x_1,x_2,M^2)=x_1x_2 \sum_i & &\, 
e_i^2\left[ q_i^A(x_1, M^2)\overline{q}_i^B(x_2,M^2) \right .\nonumber\\
& + &\left. \overline{q}_i^B(x_1,M^2)q_i^A(x_2,M^2)\right],
\end{eqnarray}
$q_i$ are quark structure functions,
and the sum runs over the quark flavours. 
We further have,
\begin{eqnarray}
x_1&= &M e^y/\sqrt{s}, \nonumber\\
x_2 & = & M e^{-y}/\sqrt{s}.
\end{eqnarray}
In the above $s$ is the square of the centre of mass energy per nucleon.
The parton cross sections are scaled by the nuclear thickness function
$T_{AB}(b=0)$ to obtain $dN/dM^2 \,dy$
for the Drell-Yan production of dileptons in central (impact parameter b=0)
A+B nuclear collisions. 
As observed in the
introduction, if the partons do not have any intrinsic momentum 
then the net $q_T$ of these lepton pairs would be identically
zero. The effect of including the intrinsic 
transverse momentum of the partons, to the LO Drell-Yan
can be seen
through the standard practice of folding-in a 
Gaussian distribution in $q_T$ with a width of $\sim$ 400 MeV.
In the following we use the CTEQ4M parametrization~\cite{cteq5} of the
parton distributions.

\begin{figure}[tb]
  \begin{center}
  \epsfig{file=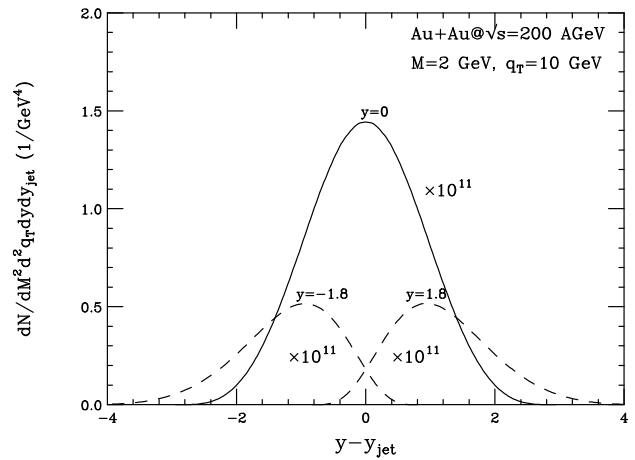,width=8.2cm}
  \caption{Rapidity correlation of the dilepton and the recoiling jet
at RHIC.}
  \label{fig2}
  \end{center}
\end{figure}

\subsection{NLO Drell-Yan :  Annihilation}

The NLO process for dilepton production has several interesting
features relevant to the present study which we recall here. 
We reiterate that we are interested only in 
large $q_T$ processes, and thus the treatment discussed here
is sufficient for this purpose \cite{berger}. A full treatment~\cite{kaufmann}  
which regularizes the behaviour of the cross section as
$q_T~\to~0$ is of course available.
First consider the annihilation process, $ q+\overline{q} \to
g+\gamma^*$. The differential cross section for the production
of a dilepton having mass $M$, transverse momentum $q_T$, rapidity $y$,
with a jet associated at the rapidity $y_{\rm jet}$ through the annihilation 
process is given by
\begin{equation}
\frac{d\sigma^{\rm A}}{dM^2 \ d^2q_T\ dy \ dy_{\rm jet}}=
F(x_1,x_2,\mu_F)\frac{1}{\pi}\frac{d\widehat{\sigma}^{A}}{dM^2 \ d\widehat{t}}.
\label{ady}
\end{equation}
We have \cite{kaj}
\begin{equation}
\frac{d\widehat{\sigma}^{A}}{dM^2 \ d\widehat{t}}
=\frac{4}{9} \, \frac{2\alpha^2 \alpha_s(\mu_R)}{3M^2} \frac{(\hat{t}-M^2)^2+
(\hat{u}-M^2)^2}{\hat{s}^2\hat{t}\hat{u}}~,
\end{equation}
where
\begin{eqnarray}
\hat{s}-M^2&=&x_1 x_2 s -M^2 \nonumber\\
\hat{t}-M^2&=&-\frac{1}{2}\ s \ \overline{x}_T x_1 e^{-y}\nonumber\\
\hat{u}-M^2&=&-\frac{1}{2}\ s  \ \overline{x}_T x_2 e^y \nonumber\\
\hat{t} \ \hat{u}&=&\hat{s} \ q_T^2
\end{eqnarray}
and
\begin{eqnarray}
x_1&=&\frac{1}{2}\overline{x}_T e^y+\frac{1}{2}x_T e^{y_{\rm jet}}\nonumber\\
x_2&=&\frac{1}{2}\overline{x}_T e^{-y} +\frac{1}{2}x_T e^{-y_{\rm jet}}.
\label{x1x2}
\end{eqnarray}

\begin{figure}[tb]
  \begin{center}
  \epsfig{file=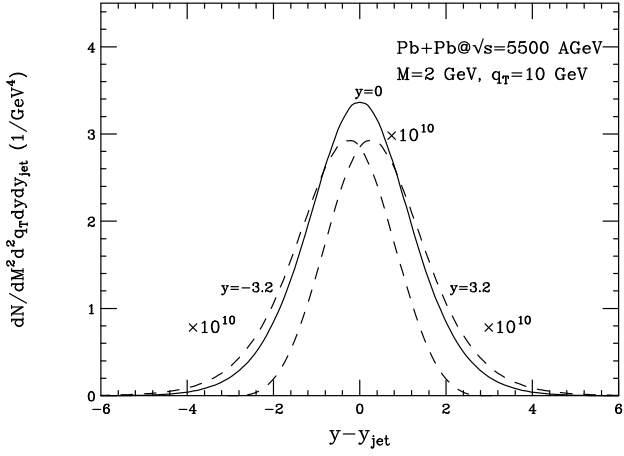,width=8.2cm}
  \caption{Rapidity correlation of the dilepton and the recoiling jet
at the LHC.}
  \label{fig3}
  \end{center}
\end{figure}

Above, any quantity under a caret (e.g. $\hat{\sigma}$) is associated with a
parton-parton process. 
We set the renormalization and the factorization scales to 
 $\mu_R=\mu_F=M^2+q_T^2$, as appropriate for the study of the transverse
momentum of the lepton-pairs. It is obvious that if one decides to
choose $\mu_F=\mu_R=M^2$, the production of NLO Drell-Yan dileptons will
change.
We further have 
\begin{equation}
\overline{x}_T^2=x_T^2+4\tau~, 
x_T=\frac{2q_T}{\sqrt{s}}~,
\tau=\frac{M^2}{s}.
\end{equation}
We also add that the momenta of the lepton pair $q$, the incoming partons
($p_i$), and the recoiling parton ($k$),
in the nucleon-nucleon centre of mass system are given by:
\begin{eqnarray}
p_1 & = & x_1 \frac{1}{2}\sqrt{s} (1,0,0,+1),\nonumber\\
p_2 & = & x_2 \frac{1}{2}\sqrt{s} (1,0,0,-1),\nonumber\\
q   & =& (q_0,{\bf q}_T,q_L),\nonumber\\
k & = & (k_0,-{\bf q}_T,k_L).
\end{eqnarray}

\begin{figure}[tb]
  \begin{center}
  \epsfig{file=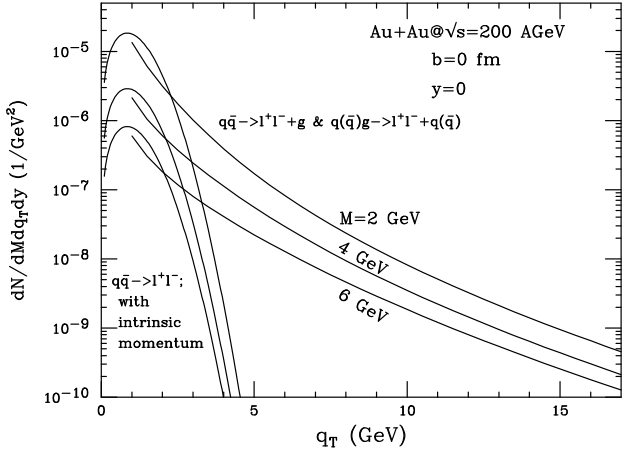,width=8.2cm}
  \caption{Jet-rapidity integrated dilepton rates at RHIC
for various masses of the pair.}
  \label{fig4}
  \end{center}
\end{figure}

\subsection{NLO Drell-Yan :  Compton}
The expression for the Compton contribution to the Drell-Yan process at NLO is
a little more involved. We have \cite{kaj}
\begin{eqnarray}
\frac{d\sigma^{\rm C}}{dM^2 \ d^2q_T\ dy \ dy_{\rm jet}}&=&
F_1(x_1,x_2)\frac{1}{\pi}\frac{d\widehat{\sigma}^{C}(\hat{s},
\hat{u})}{dM^2 \ d\widehat{u}}+ \nonumber\\
& &
F_2(x_1,x_2)\frac{1}{\pi}\frac{d\widehat{\sigma}^{C}(\hat{s},
\hat{t})}{dM^2 \ d\widehat{u}},
\label{cdy}
\end{eqnarray}
where
\begin{eqnarray}
F_1(x_1,x_2)&=&x_1 x_2 g^A(x_1) \sum_i e_i^2 
\left[q_i^B (x_2)+ \overline{q}_i^B(x_2) \right]\ , \nonumber\\
F_2(x_1,x_2)&=&x_1 x_2 \sum_i e_i^2 
\left[q_i^A (x_1)+\overline{q}_i^A (x_1) \right] g^B(x_2)\ , \nonumber\\
& &
\end{eqnarray}
\begin{equation}
\frac{d\widehat{\sigma}^{C}(\hat{s},\hat{u})}{dM^2 \ d\widehat{u}}
=\frac{1}{6} \, \frac{2\alpha^2 \alpha_s}{3M^2} \frac{(\hat{s}-M^2)^2+
(\hat{u}-M^2)^2}{\hat{s}^3(-\hat{u})},
\end{equation}
and
\begin{equation}
\frac{d\widehat{\sigma}^{C}(\hat{s},\hat{t})}{dM^2 \ d\widehat{u}}
=\frac{1}{6} \, \frac{2\alpha^2 \alpha_s}{3M^2} \frac{(\hat{s}-M^2)^2+
(\hat{t}-M^2)^2}{\hat{s}^3(-\hat{t})}.
\end{equation}
In the above, we have not explicitly shown the factorization and the
renormalization scales, which are taken as $\mu_F=\mu_R=M^2+q_T^2$
as for the NLO annihilation term.

\begin{figure}[tb]
  \begin{center}
  \epsfig{file=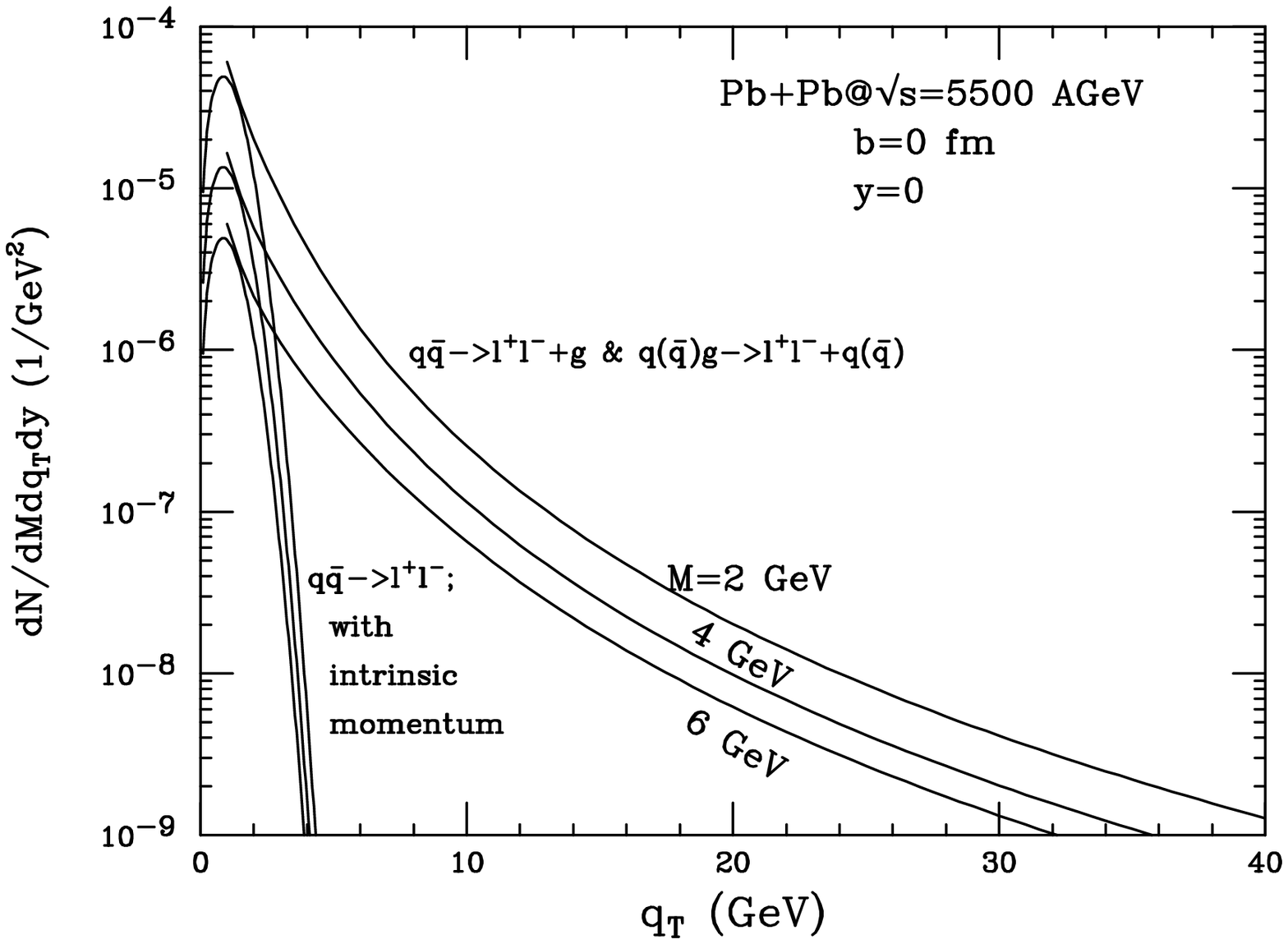,width=8.2cm}
  \caption{Jet-rapidity integrated dilepton rates at the LHC
for various masses of the pair.}
  \label{fig5}
  \end{center}
\end{figure}

\subsection{Heavy-Quark Production}
We consider the production of heavy quarks at NLO. The lowest order
contributions come from the processes
$q\overline{q} \to Q\overline{Q}$, $gg \to Q\overline{Q}$. The NLO
terms originate from $q\overline{q} \to Q\overline{Q}g$, $gq(\overline{q}) \to
Q\overline{Q} q(\overline{q})$, and $gg \to Q\overline{Q} g$. The general
expression for the production can be written as
\begin{equation}
d\sigma=\sum_{i,j} \int \ dx_1 \ dx_2 d\widehat{\sigma}_{ij}(\hat{s},M_Q^2,Q^2)
f_i^A(x_1,Q^2)
f_j^B(x_2,Q^2)
\end{equation}
where $f_i^A$ are the distribution functions for the partons in the nucleon
in the nucleus A, $x_i$ are the fractional momenta of the incoming partons,
and $\hat{s}=x_1x_2s$ is the parton-parton centre of mass energy. The parton
cross-section $\widehat{\sigma}_{i,j}(\hat{s},M_Q^2,Q^2)$ can be written as:  
\begin{equation}
\widehat{\sigma}_{i,j}(\hat{s},M_Q^2,Q^2)=\frac{\alpha_s^2(Q^2)}{M_Q^2}
f_{i,j}\left(\rho,\frac{Q^2}{M_Q^2}\right),
\end{equation}
where
\begin{equation}
\rho=\frac{4 M_Q^2}{x_1x_2 s}
\end{equation}
and
\begin{eqnarray}
f_{i,j}\left(\rho,Q^2/M_Q^2\right)=& &f_{i,j}^{(0)}(\rho)
+4\pi\alpha_s(Q^2)
\left[f_{i,j}^{(1)}(\rho)+\right.\nonumber\\
& &
\left.\overline{f}_{i,j}^{(1)}(\rho) \ln(Q^2/M_Q^2)\right],
\end{eqnarray}
 are taken from Ref.~\cite{mnr}.
We have used the factorization and the renormalization scales
as $Q^2=M_Q^2+ <p_T^2>$, where $<p_T^2>$ is the average of the
transverse momentum of the produced quark and anti-quark. 
 The correlated 
semi-leptonic decay of the heavy (D or B) mesons
is then estimated by fragmentation of the heavy-quarks to
produce the meson and then estimating its decay in the
quark model~\cite{alta}. The branching ratios to leptons for the 
decay of the D and B mesons are taken as 12\% and 
10\% ~\cite{ramona}, respectively. 
 Even though more sophisticated approaches exist \cite{isgur}, that outlined above is 
sufficient for our purpose, as we only 
wish to know where these contributions are small. We have verified that
our results are quantitatively similar to those 
obtained from PYTHIA~\cite{pythia}
and also by Gavin {\it et al.}~\cite{ramona}, where parametrizations of 
the measured lepton spectra from semi-leptonic decay of B and D mesons
are used to generate the lepton momenta.
The total cross sections for heavy $q \bar{q}$ we thus obtain are shown in Table \ref{table}.
These values do depend on the choice of structure functions and scales \cite{gavai}. 

\begin{table}[h!]
\begin{center}
\begin{tabular}{ c | c | c | c | c }
  \hline\hline
  $\sqrt{s}$ (GeV)\  & $\sigma_{\rm c \bar{c}}^{\rm LO}$ &   $\sigma_{\rm c \bar{c}}^{\rm NLO}$ & 
   $\sigma_{\rm b \bar{b}}^{\rm LO}$ &  $\sigma_{\rm b \bar{b}}^{\rm NLO}$\\ \hline
   200   & 59.3  & 145.9 & 0.84 & 1.67\\ 
   5500  & 1054 & 3362 & 79.7 & 174.4 \\
   \hline\hline
\end{tabular}
  \caption{Total cross sections for heavy $q \bar{q}$ production in pp collisions. All values 
  are in $\mu$b.} 
  \label{table}
\end{center}
\end{table}

Finally, the momentum lost by the heavy quarks during fragmentation is taken 
to be negligible \cite{vbh}.

\begin{figure}[tb]
  \begin{center}
  \epsfig{file=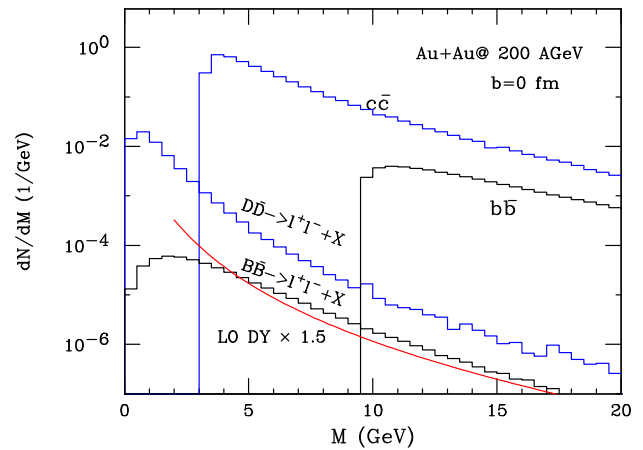,width=8.2cm}
  \caption{Mass distribution of heavy quark pairs and the dileptons from
their correlated decay at RHIC energies. The LO Drell-Yan contribution
is also shown.}
  \label{fig6}
  \end{center}
\end{figure}

\section{Results}

As a first step we explore the range of fractional momenta of the partons
(Eq.~\ref{x1x2}) which are relevant for our purpose (see Fig.~1). The 
diagonal provides the values for the situation when the rapidity of the
dilepton ($y$) as well as that of the recoiling jet ($y_{\rm jet}$) is zero.
The squares denote the point when the $y_{\rm jet}=+0.5,  $ while the
point  $y_{\rm jet}=-0.5$ is indicated by the diamonds, when the dilepton 
$y=0$. Thus large $M$ and very large $q_T$
of the lepton pair do not sample very small values of $x$, and  
the nuclear modification of the parton distribution functions will hence not 
play a large role.

\begin{figure}[tb]
  \begin{center}
  \epsfig{file=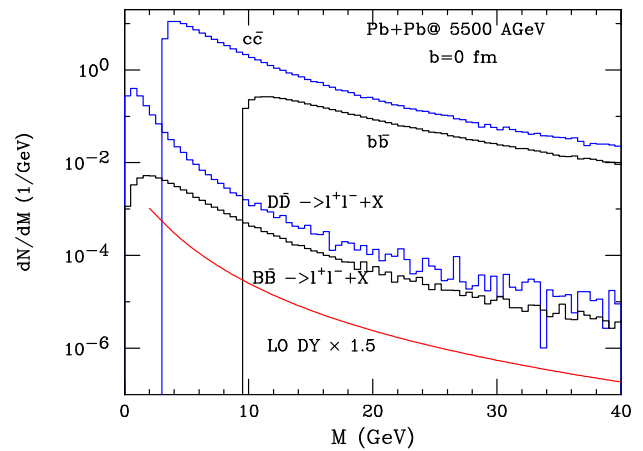,width=8.2cm}
  \caption{Same as Fig.\ref{fig6} for the LHC}
  \label{fig7}
  \end{center}
\end{figure}

Next we study the rapidity correlation between the dilepton and the jet.
We see (Figs.~2 and 3) that there is a strong positive correlation
between the rapidity of the dilepton and the recoiling jet; i.e. when the 
dilepton has a positive rapidity the jet also has a positive rapidity,
and the correlation has a width of about 2 units of rapidity. 
\begin{figure}[tb]
  \begin{center}
  \epsfig{file=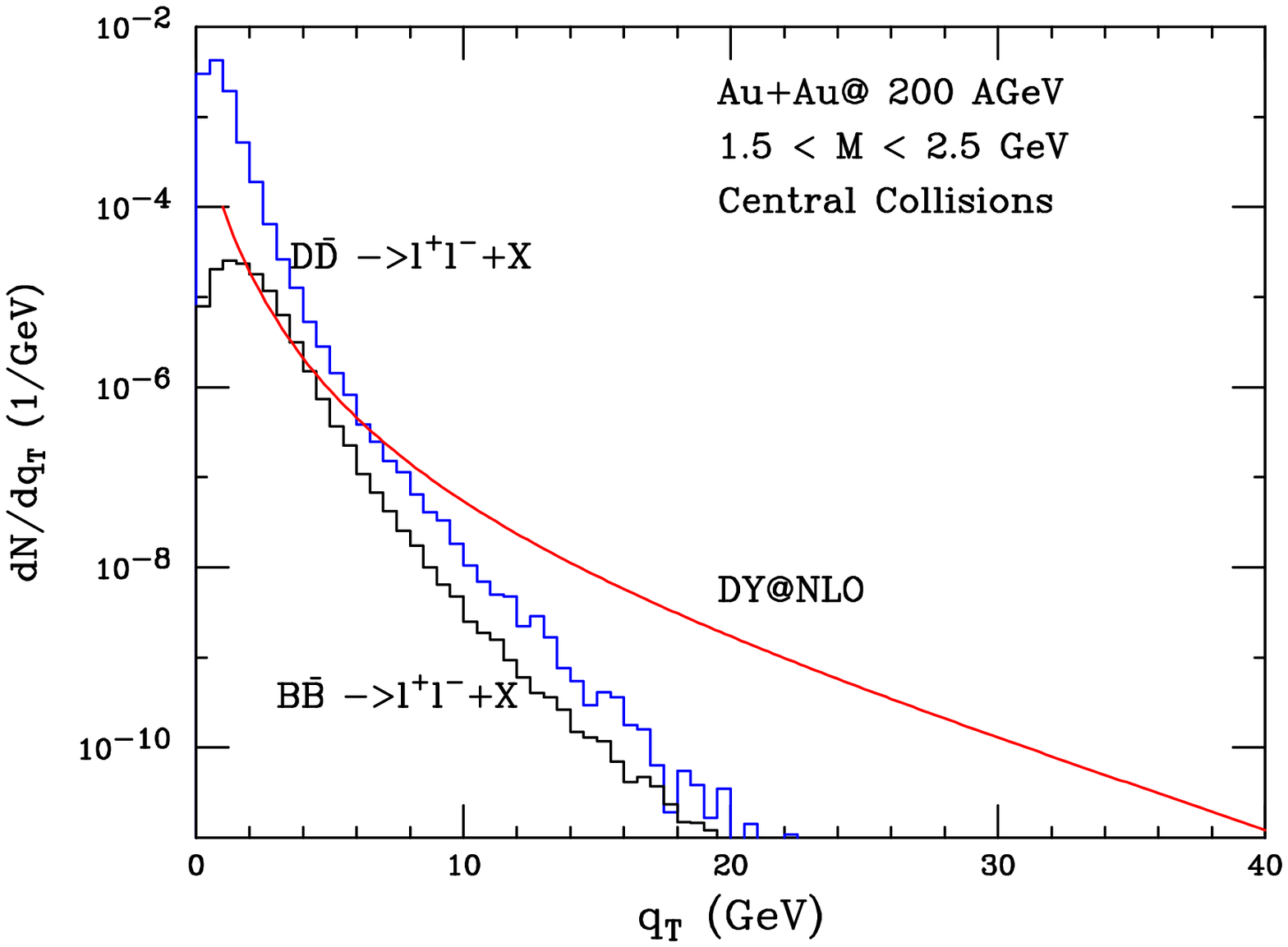,width=8.2cm}
%
  \epsfig{file=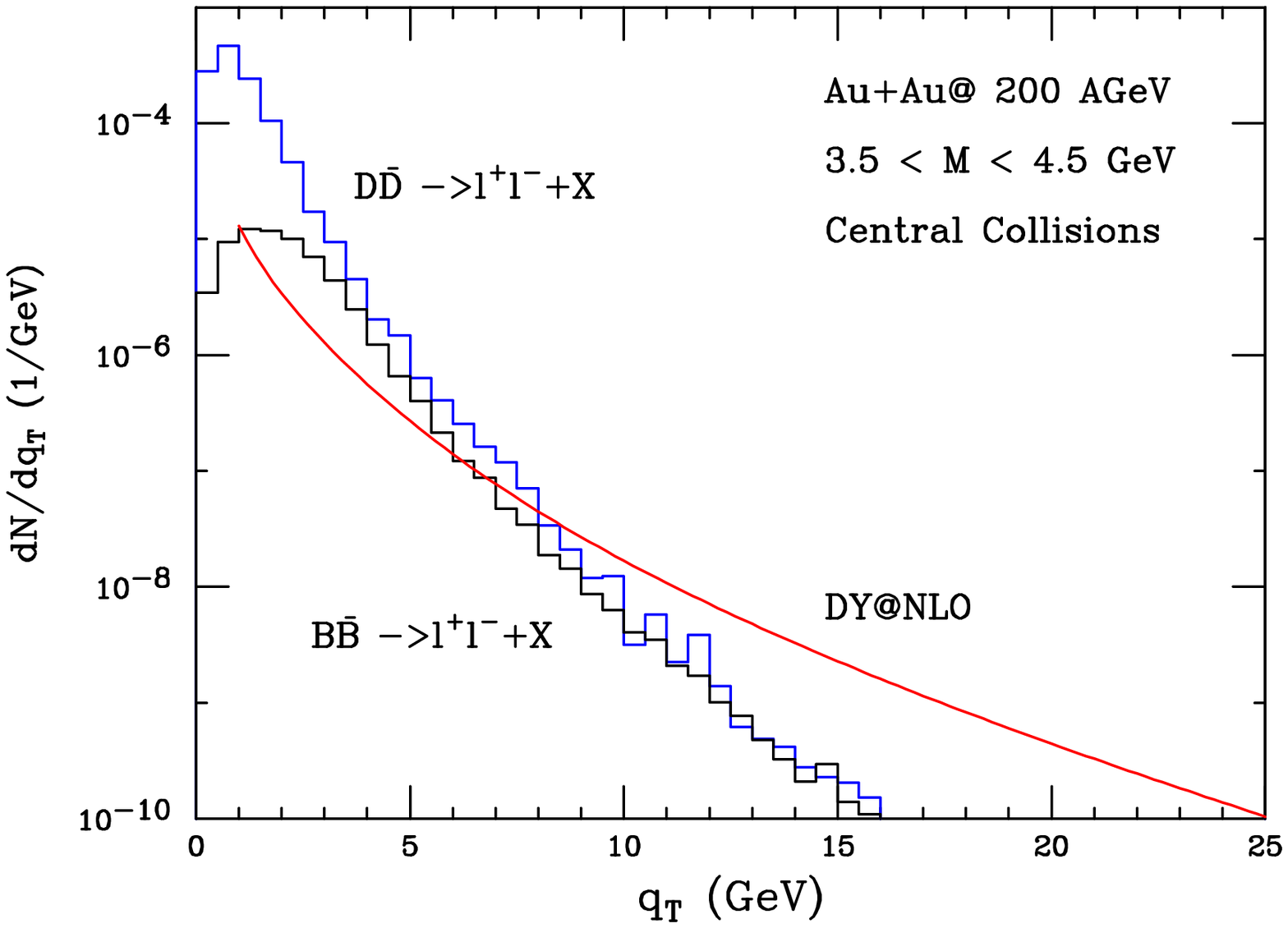,width=8.2cm}
  \caption{Rate of dilepton production with pair mass $1.5 < M < 2.5$
GeV (upper panel) and $3.5 < M < 4.5 $ GeV (lower panel) 
as a function of pair transverse momentum at RHIC energies.}
  \label{fig8}
  \end{center}
\end{figure}
\begin{figure}[tb]
  \begin{center}
  \epsfig{file=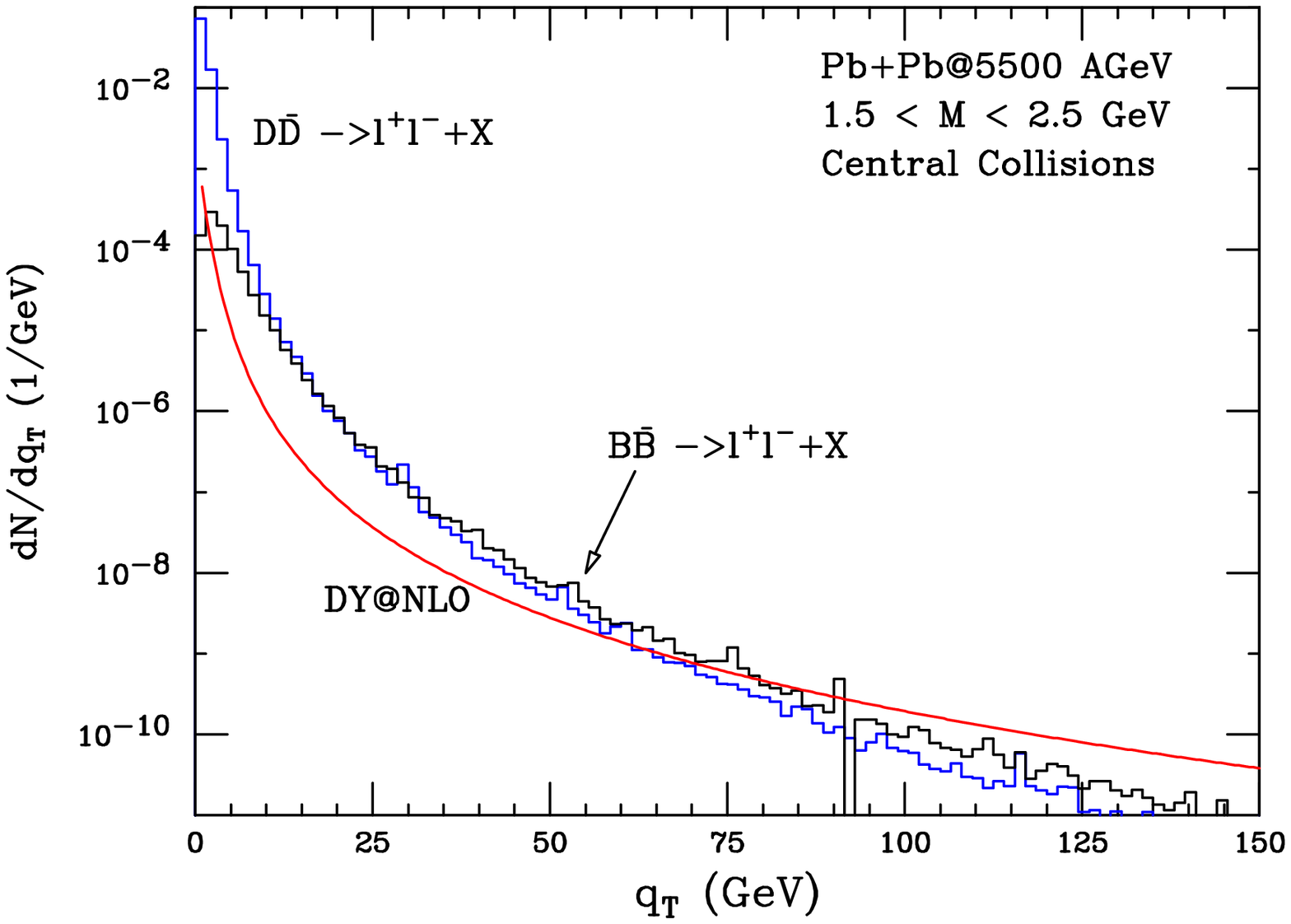,width=8.2cm}
%
  \epsfig{file=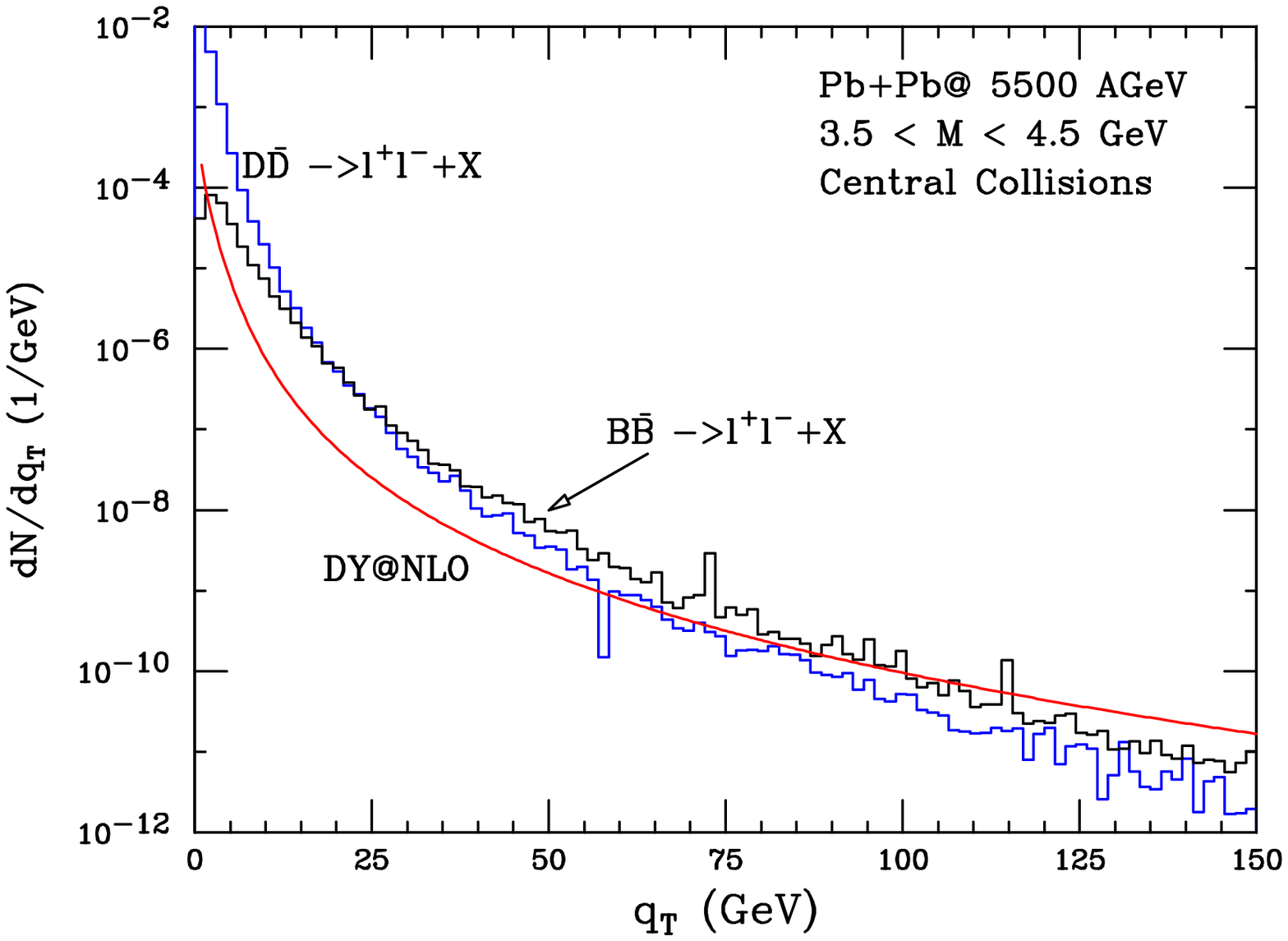,width=8.2cm}
  \caption{Same as Fig.\ref{fig8} for the LHC. }
  \label{fig9}
  \end{center}
\end{figure}
This
aspect has been discussed in detail by Kajantie {\it et al.}~\cite{kaj}
and has been attributed to the dominance of the Compton
term through the $1/\widehat{s}$ term in the scattering cross section.
Note that the finite $y$ values 
are chosen as the mid-points of the forward/backward rapidity
coverage of PHENIX at RHIC, and they fall within those for the
ALICE and 
CMS experiments at the LHC.

We now study the jet-rapidity integrated dilepton rates at RHIC (Fig.~4)
and at the LHC (Fig.~5) due to the NLO processes discussed above. We have also
given the results for the LO Drell-Yan contribution, supplemented with
a modest intrinsic $k_T$ for the partons. As expected, 
the NLO processes dominate as the transverse momentum of the 
pair (which is also the transverse momentum of the recoiling jet) increases.
The NLO contributions to the Drell-Yan and also the heavy-quark
production (see below) have been calculated ignoring the 
intrinsic momenta of the partons. Including it will increase the 
$<q_T>$ of the resulting lepton-pair and may enhance the yield by
50--100\% depending on the value of the intrinsic momentum.

The semi-leptonic decay of D and B mesons gives a large contribution to the
dilepton production \cite{ramona}. We show the results of our 
calculations, which extend up to large lepton pair masses.
The LO Drell-Yan contribution has been scaled by a factor of 1.5
to account for the NLO terms as 
our exploratory calculations will be inadequate at low $q_T$,
and a momentum integral has to be performed in order to obtain the 
mass distributions.

Figs.~\ref{fig6}~ and ~ \ref{fig7} suggest that it may 
be very difficult to observe the dileptons recoiling against a jet in the
background of the correlated decay of charm and bottom. This indeed is
the case for dileptons radiated from thermal sources~\cite{ramona}.
However, this conclusion changes when
the $q_T$ distribution of the dileptons is studied for large pair masses,
as shown in  Fig.~\ref{fig8} for mass windows of 2 and 4
GeV at RHIC energies. We find that dileptons 
originating from the NLO Drell-Yan process dominate at large $q_T$,
and with anticipated total sampled event sizes of
$\sim 10^{10}$ events, 
open a window for a clean measurement of dilepton-tagged-jets.
This window shifts to higher $q_T$ at the LHC (Fig.\ref{fig9}), and 
identification becomes more difficult than at RHIC. 
The situation may improve if the heavy-quarks lose energy 
in the medium, degrading 
the D and B meson decay contributions to lower dilepton $q_T$.
Furthermore, the background can be reduced if the
contributions of the heavy-meson decays can be isolated and rejected, 
such as by determination of the decay vertices, or by jet-like isolation cuts.

\section{Conclusions}
In brief, we have estimated the production of jets recoiling against a
dilepton produced in the Drell-Yan process at NLO in relativistic heavy ion 
collisions at RHIC and LHC energies. The dileptons could be used to 
tag the jet and give the energy of the progenitor parton, which would 
provide precise information with which to determine the rate of energy
loss of partons during their passage through quark gluon plasma.
The background from the correlated decay of charm and bottom mesons, which 
plagues the identification of thermal dileptons~\cite{ramona}, is found
to be unimportant for large jet (or dilepton) transverse momenta, that is,
in the kinematical region of interest for jet-quenching studies.
Even though the
counting rates will be small, the anticipated event samples suggest that
dilepton jet-tagging is feasible. 
 This method
could  thus prove useful in the resolution of several issues of fundamental 
importance in the physics of high energy partons
traversing a plasma of quarks and gluons.

\acknowledgments  

We thank M. Mangano for providing us
with his code for heavy-quark production \cite{mnr} and
for many helpful communications. 
This work was supported by the Natural Sciences and Engineering Research
Council of Canada. ORNL is managed by UT-Battelle, LLC, for the U.S. 
Department of Energy under contract DE-AC05-00OR22725.

\end{document}